\documentclass[useAMS,usenatbib]{mn2e}
\usepackage{epsfig}

\title[kHz QPOs in Aql X-1]{Discovery of the upper kilo-Hz QPO from the X-ray transient Aql X-1}
\author[Didier Barret, Martin Boutelier,  M. Coleman Miller]{Didier Barret$^{1}$\thanks{E-mail:
Didier.Barret@cesr.fr}, Martin Boutelier$^{1}$, M. Coleman Miller$^{2}$ \\
$^{1}$Centre d' Etude Spatiale des Rayonnements, CNRS/UPS, 9 Avenue du Colonel Roche, 31028 Toulouse Cedex04, France \\
$^{2}$Department of Astronomy, University of Maryland, College Park, MD 20742-2421, United States }
\begin{document}

\date{}

\pagerange{\pageref{firstpage}-\pageref{lastpage}} \pubyear{2007}

\maketitle

\label{firstpage}

\begin{abstract}
We report on a comprehensive analysis of the kilo-Hz ($\ge 600$ Hz)
quasi-periodic oscillations (kHz QPOs) detected from the neutron star
X-ray transient Aquila X-1 (Aql X-1) with the Rossi X-ray Timing Explorer,
between 1997 and 2007. Among kHz QPO sources, Aql X-1 is
peculiar because so far only one kHz QPO has been reported,
whereas in most sources, two kHz QPOs are usually detected (a
lower and an upper kHz QPO). The identification of the QPOs
reported so far has therefore been ambiguous, although it has been proposed
that they were likely to be the lower QPO. Following up on previous
work, we confirm the identification of the QPOs previously reported
as lower QPOs, because of their high quality factors and the
quality factor versus frequency dependency, which are similar to those
observed in other sources. Combining all segments of data
containing a lower QPO, we detect for the first time an upper
kHz QPO. As in other sources for which the neutron star spin
frequency is larger than 400 Hz (550.25 Hz in Aql X-1), the
frequency difference between the two kHz QPOs is close to half
the spin frequency. Based on this result, we re-examine the link between the
neutron star spin and the frequency of the kHz QPOs, to show that a model 
in which the separation of the lower and upper QPOs relates to the neutron 
star spin frequency is still as good as any comparably simple model.

\end{abstract}


\section{Introduction}
Since its launch in 1995, the Rossi X-ray Timing Explorer (Bradt,
Swank and Rothschild, 1993) has detected kHz QPOs in more than
25 accreting systems containing a weakly magnetized neutron star
(see van der Klis 2006 for a review). In most sources, two kHz QPOs
are usually detected; the lower of the two is seen as a
relatively narrow peak in the Fourier Power Density Spectrum (PDS)
with a quality factor ($Q=\nu/\Delta\nu$) exceeding 200 (e.g., Barret, et al. 2005a), whereas the
upper QPO is typically much broader ($Q\sim 5-20$). This makes the
lower QPO easier to detect, especially on short timescales.
Both the lower and upper QPOs vary in frequency with time, but
the frequency difference remains always close to the spin
frequency of the neutron star (or half its value). Removing the contribution of the frequency drift to the
measured QPO width, Barret, Olive, Miller (2005) have
demonstrated that the lower and upper QPOs follow a different path
in a quality factor versus frequency diagram. In particular, they
found that the quality factor of the lower kHz QPOs increases
smoothly with frequency, saturates at a maximum value, beyond which a sharp
drop is observed (Barret, Olive, Miller 2005). The same behaviour has been observed in several
different systems (Barret, Olive, Miller 2006), suggesting that the drop is related to a special
location in spacetime, e.g., the innermost stable circular orbit (ISCO) (Barret, Olive, Miller, 2006, 2007).

Previous investigations of the kHz QPOs detected from the recurrent X-ray
transient Aql X-1 (Zhang et al. 1998, Cui et al. 1998, Reig et al.
2000, M\'endez et al. 2001, Reig et al. 2004) have reported only a single QPO, making
its identification difficult. By comparison with the properties of
lower QPOs seen in twin QPO sources (correlation of the
frequency and spectral colors, QPO width, RMS-frequency
dependency, energy spectrum of the QPOs), it was however
proposed that the QPOs detected were likely to be lower QPOs (M\'endez et al. 2001).  

Aql X-1 contains a rapidly rotating
neutron star, spinning at 550.25 Hz, as inferred from the
discovery of an episode of coherent pulsation in its persistent
emission (Casella et al. 2007).  This frequency is close to the
previously detected frequency of X-ray burst oscillations (Zhang
et al. 1998). By analogy with other sources spinning at a
frequency above 400 Hz, one would therefore expect an upper QPO
to be detected with a frequency separation close to 275 Hz (see van der Klis 2006 for a review). Despite
extensive searches, no such QPO has yet been reported.

In this paper, we analyze in a homogenous way all archival RXTE data on Aql X-1, with the goal of studying the
quality factor of the QPOs and searching for an upper QPO. The data used here have been presented in Zhang et al. (1998), Cui et al. (1998), Reig et al.
(2000), M\'endez et al. (2001) and Reig et al. (2004). In the next section, we present our analysis scheme, which builds upon previous previously described procedures in Barret, Olive \& Miller (2005, 2006, 2007).  We then present our results, which reveal for the first time 1) that the quality factor of the lower QPOs follows the same pattern seen in other sources (e.g. 4U~1608--522 \& 4U~1636--536, Barret et al. 2006), and 2) Aql X-1 displays an upper QPO that is detected close to half the spin frequency of the neutron star (275 Hz), when combining all the data available. Starting with this result, we then discuss on the link between kHz QPOs and neutron star spin.

\begin{table*}
 \centering
   \caption{QPOs from Aql X-1. The observations are ObsID sorted. The name of the ObsID, the starting date of the observation, the cumulative integration time of all the 16 PDS shifted (T$_{obs}$, the fraction of PDS per ObsID, for which an instantaneous frequency could be estimated varies from about 40\% up to 100 \%), the total source count rate (Cts/s), the QPO frequency  ($\bar{\nu}$) to which all the PDS were shifted, the range of variation of the QPO frequency within the ObsID ($\nu_{min}-\nu_{max}$), the mean quality ($\bar{Q}$) and mean amplitude (RMS) are listed. All errors are computed such that $\Delta\chi^2=1$. The RMS is computed as $\sqrt{I_{lor}/S}$, where $I_{lor}$ is the fitted Lorentzian integral and $S$ is the source count rate. Because the error on the source count rate is negligible, the error on RMS is computed as $1/2 \times RMS \times \Delta I_{lor}/I_{lor}$, where $ \Delta I_{lor}$ is the error on $I_{lor}$, computed with $\Delta\chi^2=1$.  The significance of the QPO which is $I_{lor}/\Delta I_{lor}$ can thus be retreived from the error quoted on the RMS: it exceeds typically $\sim 7\sigma$ and goes up to $\sim 20\sigma$. The high quality factor recovered suggests that the QPOs detected are all lower kHz QPOs.  Previous analysis of these data have been presented in Zhang et al. (1998), Cui et al. (1998), Reig et al. (2000), M\'endez et al. (2001) and Reig et al. (2004).}
  \begin{tabular}{@{}|cccccccc|@{}}
\hline 
 ObsID & Date & T$_{obs}$ & Cts/s & $\bar{\nu}$ & $\nu_{min}-\nu_{max}$ &$\bar{Q}$ & RMS (\%)   \\
  \hline
20092-01-01-02 & 1997/08/13-11:09:35 &  656.0 & 1380.2 &  833.7 & $ 829.0- 837.6$ & $ 129.7\pm  14.7$ & $ 7.2\pm 0.3$ \\
20092-01-02-01 & 1997/08/15-19:03:17 & 1232.0 & 1520.1 &  875.2 & $ 869.5- 883.0$ & $  96.8\pm  15.3$ & $ 5.6\pm 0.3$ \\
20092-01-02-03 & 1997/08/17-06:06:08 &  592.0 & 1571.9 &  848.3 & $ 846.3- 850.6$ & $ 115.3\pm  16.8$ & $ 6.6\pm 0.4$ \\
20092-01-05-01 & 1997/09/06-12:50:24 & 1360.0 & 1313.1 &  877.8 & $ 874.3- 885.5$ & $ 165.0\pm  25.1$ & $ 5.6\pm 0.3$ \\
20092-01-05-01 & 1997/09/06-14:14:55 & 1504.0 & 1380.3 &  886.2 & $ 883.1- 892.2$ & $ 157.0\pm  27.6$ & $ 5.2\pm 0.3$ \\
20092-01-05-01 & 1997/09/06-15:52:47 &  864.0 & 1360.4 &  890.9 & $ 885.5- 896.6$ & $ 104.4\pm  22.7$ & $ 5.5\pm 0.4$ \\
20098-03-07-00 & 1997/02/27-05:54:24 & 1040.0 & 1098.4 &  773.4 & $ 761.9- 787.5$ & $ 150.6\pm  15.8$ & $ 7.8\pm 0.3$ \\
20098-03-07-00 & 1997/02/27-06:54:23 & 3152.0 & 1111.8 &  801.9 & $ 766.7- 832.6$ & $ 129.3\pm   7.8$ & $ 8.1\pm 0.2$ \\
20098-03-08-00 & 1997/03/01-21:34:47 & 2192.0 &  778.4 &  783.0 & $ 749.0- 802.1$ & $ 102.2\pm  10.8$ & $ 8.9\pm 0.4$ \\
20098-03-08-00 & 1997/03/01-22:56:32 & 3088.0 &  765.0 &  797.9 & $ 769.5- 821.5$ & $ 111.2\pm   9.9$ & $ 9.3\pm 0.3$ \\
30072-01-01-01 & 1998/03/03-14:01:46 & 1104.0 & 1603.2 &  678.2 & $ 673.0- 684.1$ & $ 108.1\pm   9.0$ & $ 7.0\pm 0.2$ \\
30072-01-01-02 & 1998/03/04-14:00:40 & 1232.0 & 1994.6 &  856.3 & $ 849.1- 867.5$ & $ 115.5\pm  12.4$ & $ 5.8\pm 0.2$ \\
30072-01-01-03 & 1998/03/05-12:24:38 & 1104.0 & 2212.4 &  838.3 & $ 833.6- 846.1$ & $ 138.2\pm  11.4$ & $ 5.9\pm 0.2$ \\
40047-02-05-00 & 1999/05/31-15:30:26 & 2640.0 & 1112.4 &  852.7 & $ 843.9- 857.8$ & $ 131.9\pm  13.1$ & $ 6.2\pm 0.2$ \\
40047-03-02-00 & 1999/06/03-15:33:09 & 2640.0 &  975.0 &  861.1 & $ 852.0- 874.7$ & $ 158.4\pm  16.2$ & $ 6.8\pm 0.2$ \\
40047-03-03-00 & 1999/06/04-13:48:20 & 3152.0 & 1035.7 &  845.2 & $ 834.9- 859.9$ & $ 141.0\pm  11.2$ & $ 7.4\pm 0.2$ \\
40047-03-03-00 & 1999/06/04-15:23:28 & 3024.0 & 1032.2 &  862.8 & $ 851.6- 873.6$ & $ 141.4\pm  12.2$ & $ 7.3\pm 0.2$ \\
50049-02-13-00 & 2000/11/07-07:01:35 & 1952.0 & 1750.6 &  833.3 & $ 823.8- 839.5$ & $ 141.0\pm  19.3$ & $ 4.4\pm 0.2$ \\
50049-02-15-03 & 2000/11/13-20:57:38 & 2064.0 & 1164.4 &  870.3 & $ 867.3- 874.9$ & $ 136.8\pm  16.1$ & $ 6.2\pm 0.3$ \\
50049-02-15-03 & 2000/11/13-22:14:40 & 3024.0 & 1121.6 &  834.9 & $ 822.1- 848.5$ & $ 158.8\pm  11.7$ & $ 7.1\pm 0.2$ \\
50049-02-15-04 & 2000/11/14-01:42:39 & 1744.0 &  836.5 &  811.7 & $ 808.3- 815.9$ & $ 137.3\pm  17.4$ & $ 7.3\pm 0.3$ \\
50049-02-15-04 & 2000/11/14-03:02:23 & 2704.0 &  785.3 &  754.3 & $ 734.8- 767.1$ & $ 132.8\pm  13.2$ & $ 7.4\pm 0.3$ \\
50049-02-15-05 & 2000/11/15-04:30:45 & 3152.0 &  841.3 &  631.3 & $ 613.7- 642.9$ & $  84.9\pm   8.8$ & $ 7.6\pm 0.3$ \\
50049-02-15-05 & 2000/11/15-06:06:40 & 2784.0 &  825.6 &  643.8 & $ 627.7- 656.8$ & $  82.9\pm   9.4$ & $ 7.4\pm 0.3$ \\
50049-02-15-06 & 2000/11/16-04:47:43 & 1808.0 &  745.0 &  717.6 & $ 712.0- 723.6$ & $ 117.0\pm  11.8$ & $ 8.7\pm 0.3$ \\
50049-02-15-07 & 2000/11/16-06:33:43 & 1104.0 &  564.7 &  698.6 & $ 686.0- 709.1$ & $ 131.5\pm  26.9$ & $ 8.1\pm 0.6$ \\
60054-02-03-03 & 2001/07/10-18:59:27 & 1552.0 &  452.2 &  770.4 & $ 755.0- 787.2$ & $ 129.9\pm  19.0$ & $11.4\pm 0.6$ \\
60054-02-03-05 & 2001/07/12-17:15:12 & 1232.0 &  579.7 &  801.9 & $ 798.8- 813.3$ & $ 177.3\pm  28.5$ & $ 8.9\pm 0.5$ \\
60429-01-05-00 & 2002/02/18-22:14:08 & 1824.0 &  558.0 &  787.9 & $ 769.7- 819.6$ & $ 111.3\pm  16.1$ & $ 9.4\pm 0.5$ \\
60429-01-09-00 & 2002/02/27-21:42:07 & 1472.0 &  775.3 &  843.1 & $ 835.6- 850.2$ & $ 121.8\pm  20.7$ & $ 6.9\pm 0.4$ \\
70069-03-01-01 & 2002/03/07-10:26:39 & 2192.0 &  945.8 &  807.0 & $ 786.1- 828.4$ & $ 183.5\pm  16.0$ & $ 7.1\pm 0.2$ \\
70069-03-01-02 & 2002/03/07-15:17:35 & 1296.0 &  941.0 &  781.8 & $ 771.6- 797.2$ & $ 153.0\pm  18.4$ & $ 7.2\pm 0.3$ \\
70069-03-02-00 & 2002/03/11-22:17:20 & 1648.0 &  815.4 &  627.7 & $ 607.4- 637.3$ & $  83.0\pm  14.1$ & $ 7.1\pm 0.4$ \\
70069-03-02-01 & 2002/03/10-00:54:39 &  656.0 & 1593.4 &  768.5 & $ 763.4- 775.9$ & $ 169.8\pm  17.3$ & $ 7.3\pm 0.3$ \\
70069-03-03-03 & 2002/03/17-20:48:59 & 1120.0 &  859.5 &  876.6 & $ 870.2- 886.3$ & $  95.6\pm  18.5$ & $ 7.2\pm 0.5$ \\
70069-03-03-06 & 2002/03/18-12:45:43 &  656.0 &  819.6 &  821.1 & $ 813.4- 832.4$ & $ 103.1\pm  19.3$ & $ 8.3\pm 0.6$ \\
70069-03-03-07 & 2002/03/18-20:34:39 & 2336.0 &  797.1 &  827.0 & $ 816.6- 844.3$ & $ 112.6\pm  10.4$ & $ 8.5\pm 0.3$ \\
70069-03-03-09 & 2002/03/19-14:07:43 &  976.0 &  882.4 &  700.9 & $ 692.0- 711.7$ & $ 112.7\pm  15.0$ & $ 8.8\pm 0.4$ \\
70069-03-03-14 & 2002/03/21-20:03:39 & 1952.0 &  600.5 &  764.7 & $ 756.6- 771.8$ & $ 119.9\pm  15.2$ & $ 9.4\pm 0.4$ \\
  \hline
\end{tabular}
\label{dbarret_tab1}
\end{table*}

\section{Observations and results}
For the purposes of this paper, we have retrieved all science event
files for Aql X-1 from the RXTE HEASARC archives. Data, up to the July 2007 are used. The files
are identified with their observation identifier (ObsID) following the
RXTE convention. An ObsID identifies a temporally contiguous
collection of data from a single pointing. Type I X-ray bursts and
data gaps are removed from the files. 

For each file identified with its ObsID, we have computed Leahy
normalized Fourier PDS between 1 and 2048 Hz
over 16~s intervals (with 1 Hz resolution), using events with energy
between 2 and 40 keV. $N$ 16-second PDS are thus computed. $N$ is
typically around 150-200 in most files, whose duration $\sim 3000$ seconds is
consistent with the orbital period of the RXTE spacecraft. A PDS averaging the $N$ PDS  is first computed. This averaged
PDS is then searched for a high frequency QPO between 500 and 1500 Hz,
using a scanning technique which looks for peak excesses above the 
Poisson counting noise level (Boirin et al. 2000). The strongest
excess is then fitted within a 400 Hz window (200 Hz on each side of
the peak) with a Lorentzian of three parameters (frequency, full width
at half maximum, and its integral) to which a constant is added to
account  for the counting noise level (close to 2.0 in a Leahy
normalized PDS). If the significance of the fitted excess is less than
$3\sigma$, the ObsID is not considered for further processing.  A QPO 
(with $Q>10$) is detected in 47 ObsIDs. 

\begin{figure}
\includegraphics[height=0.3\textheight]{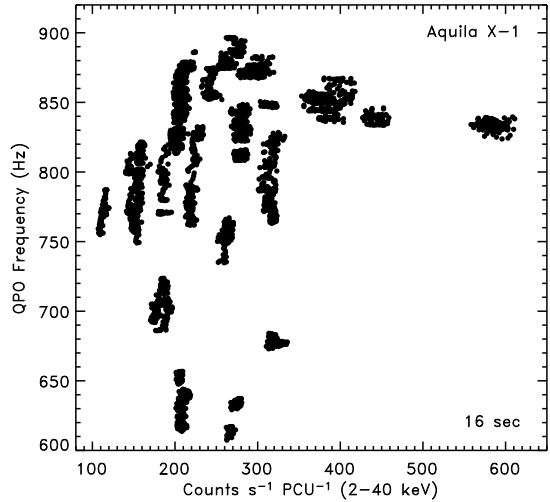}
\caption{QPO frequency versus 2-40 keV count rate. There are 4364 individual measurements, representing the count rate integrated over 16 seconds.}
\label{dbarret_fig1}
\end{figure}

We wish to estimate the quality factor of the QPO, after removing as
much as possible the contribution from the frequency drift to the
measured width (Barret et al. 2005, 2006). For this, we must first
reconstruct the time evolution of the QPO frequency, within each
ObsID of interest. Because the frequency may change significantly
(up to tens of Hz in one thousand seconds), we must consider the
shortest timescales to track those frequency changes. Given the
strength of the QPOs of Aql X-1, we have found that 256 seconds was
an appropriate timescale, which allows a homogeneous study of all
its QPOs. Using a sliding time window of 256 seconds with a time
step of 256/4=64 seconds, we have averaged 16 16-second PDS. This
PDS is then searched for an excess around the mean QPO frequency,
and the strongest excess is again fitted with a Lorentzian within a
400 Hz frequency window. Between two consecutive QPO detections
(above a given significance threshold), a linear interpolation
enables us to estimate the instantaneous QPO frequency in the
16-second PDS. In addition, the QPO may not be detected all the time
(due for instance to statistical fluctuations or a rapid frequency
jump), so gaps of duration shorter than 256 seconds are filled with
a linear interpolation. Those PDS for which no estimate of the QPO
frequency has been obtained are removed from the subsequent
analysis. Taking $3\sigma$ for the significance threshold, we were
able to reconstruct the time evolution of the QPO frequency in 39
out of the 47 ObsIDs in which a QPO was detected.
Figure~\ref{dbarret_fig1} shows the instantaneous QPO frequency
against the 2-40 keV count rate per PCA unit. As can be seen, those
QPOs are detected between 600 and 900 Hz. The source displays the
so-called parallel tracks on the left hand side of the figure, but
the tracks seem to have collapsed at higher count rates. It is
interesting to note that the high count rate high frequency part of
the diagram has not been sampled yet (unlike other sources, e.g.
4U~1636--536, Barret et al. 2005).

\begin{figure}
\includegraphics[height=0.3\textheight]{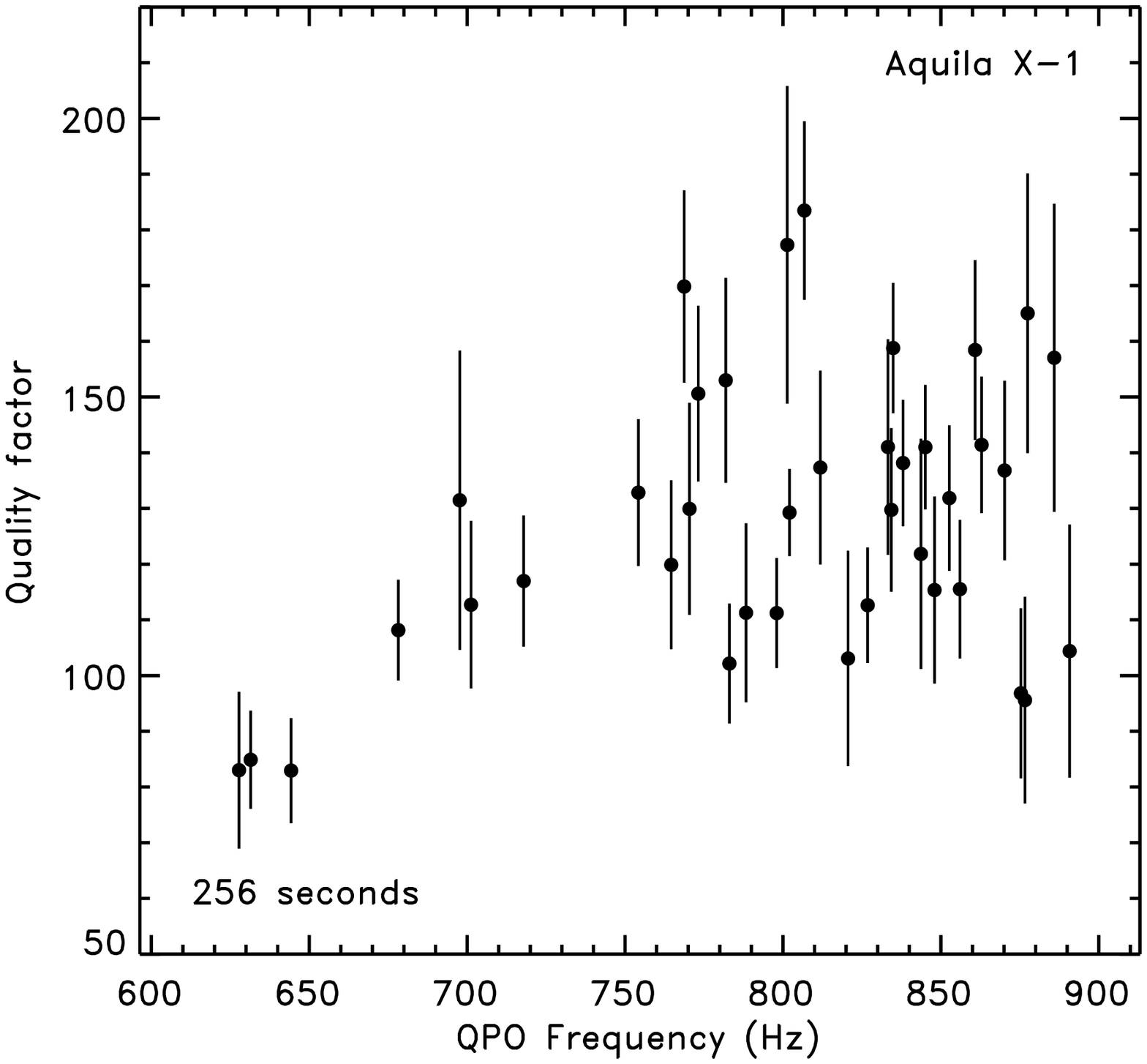}
\caption{Quality factor versus frequency dependency of the strong QPO detected for Aql X-1, as recovered within 39 ObsIDs, after correction for the frequency drift. There is a trend for Q to increase with frequency. Albeit with large scatter, a saturation in Q is also suggested. This behavior is typical of a lower QPO.}
\label{dbarret_fig2}
\end{figure}

Having reconstructed the time history of the QPO frequency, in each
ObsID, we can now shift-and-add the PDS associated with a frequency
to a reference frequency and fit the resulting QPO. In table
\ref{dbarret_tab1}, we list the measured parameters of the QPOs, in
particular its mean quality factor. The minimum Q value is 60 with
a maximum around 200. As shown in Figure~\ref{dbarret_fig2}, there
is a trend for the quality factor to increase with frequency. Both
the high Q value and its dependency suggest that these QPOs are
lower QPOs.

\begin{figure}
\includegraphics[height=0.3\textheight]{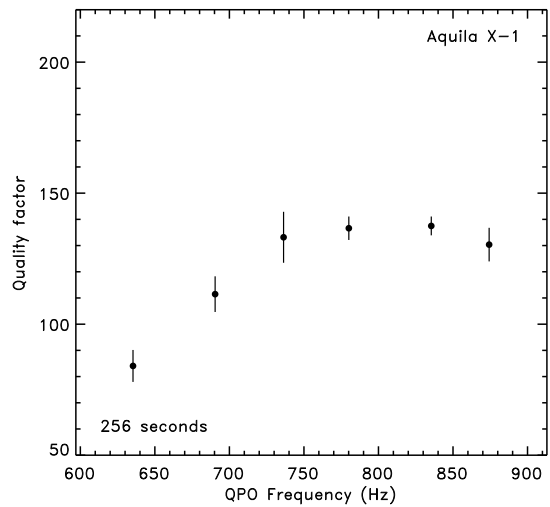}
\caption{Mean quality factor versus frequency dependency for Aql X-1. The data shown in Fig. \ref{dbarret_fig2} have been grouped over a 50 Hz bin (the same scales as Fig. 2 are used on each axis). The rise and the saturation (possibly a drop) of the quality factor is a behavior similar to that of other known lower kHz QPOs (see Barret, Olive, Miller 2006).}
\label{dbarret_fig3}
\end{figure}

In order to get a better description of the quality factor of the QPOs, we have grouped all the instantaneous frequencies with a bin of 50 Hz. All 16 second PDS falling into the same bin, are then shifted to the mean frequency and added. The mean quality factor of the QPOs so recovered is shown in Figure~\ref{dbarret_fig3}. Although the sample of QPOs is relatively limited (tens of ObsID containing a QPO, as opposed to more than two hundreds in the case of 4U1636-536, Barret et al. 2005), our data are consistent with a saturation of the quality factor, and even suggestive of a decrease at high frequencies.  This behaviour is clearly reminiscent of the lower kHz QPO, as the upper QPO is generally characterized in this frequency range by a much lower Q (up to 20 at most), rising steadily with frequency. Therefore our results support previous claims that the strong QPOs detected so far from Aql X-1 are lower QPOs (M\'endez et al., 2001) .

In addition to the observations listed in Table \ref{dbarret_tab1}, there are 8 ObsIDs in which the QPO frequency could not be tracked with the above method (shorter observations or weaker signal). For those observations (8 in total), only the mean QPO parameters over the ObsID have been measured. In table \ref{dbarret_tab2}, we list the parameters of the additional QPOs detected. Their significance is typically around $4\sigma$. Their Q values, although not corrected for the drift, are rather high ($\ge 50$, in all but one case) suggesting that they may also be lower QPOs. The highest frequency measured is about 900 Hz. It is unfortunate that we have not been able to measure the quality factor at this frequency. If the drop seen in other sources is also present in Aql X-1, one would expect the quality factor to be less than the one where it saturates. 

\begin{table*}
 \centering
   \caption{QPOs from Aql X-1 for which the correction for the frequency drift could not be applied, when estimating the quality factor. As expected, the QPOs have lower Q factors on average than the one listed in Table \ref{dbarret_tab1}. The name of the ObsID, the starting date of the observation, the PDS integration time (T$_{obs}$), the total source count rate, the mean QPO frequency ($\bar{\nu}$), the mean quality factor ($\bar{Q}$), the mean amplitude (RMS), the significance of the detection ($\sigma$) are listed. All errors are computed such that $\Delta\chi^2=1$. Although uncorrected for the frequency drift, the high quality factor of those QPOs suggests that they are also lower QPOs.}
  \begin{tabular}{@{}|cccccccccc|@{}}
\hline 
 ObsID & Date & T$_{obs}$ & Cts/s & $\bar{\nu}$ &$\bar{Q}$  & RMS (\%) & $\sigma$ \\
  \hline
20092-01-01-01 & 1997/08/13-00:05:01 &  416.0 & 1419.6 & $ 855.6\pm   0.8$ & $  94.6\pm  25.7$ & $ 6.2\pm 0.6$ & 5.3\\
20092-01-04-02 & 1997/09/03-01:38:02 & 1472.0 & 1541.0 & $ 903.2\pm   1.9$ & $  76.8\pm  25.3$ & $ 3.7\pm 0.5$ & 3.8\\
30188-03-01-00 & 1998/03/06-08:07:12 & 1472.0 & 2442.1 & $ 870.4\pm   1.3$ & $  59.5\pm  13.5$ & $ 3.8\pm 0.3$ & 5.8\\
40047-02-05-00 & 1999/05/31-18:48:32 & 2800.0 & 1521.3 & $ 893.7\pm   2.6$ & $  44.2\pm  18.3$ & $ 3.7\pm 0.6$ & 3.4\\
40047-03-02-00 & 1999/06/03-17:02:23 & 3264.0 &  991.3 & $ 892.0\pm   1.3$ & $  61.0\pm  11.0$ & $ 5.5\pm 0.4$ & 6.7\\
40047-03-02-00 & 1999/06/03-18:41:19 & 2896.0 & 1002.9 & $ 896.7\pm   1.6$ & $  62.9\pm  22.5$ & $ 4.5\pm 0.6$ & 3.9\\
50049-02-15-03C & 2000/11/13-20:36:31 & 1136.0 & 1189.4 & $ 877.1\pm   2.4$ & $  76.6\pm  30.1$ & $ 4.5\pm 0.6$ & 3.5\\
50049-02-15-05 & 2000/11/15-07:43:28 & 3152.0 &  801.3 & $ 600.9\pm   4.4$ & $  13.0\pm   4.2$ & $ 7.6\pm 1.0$ & 3.8\\
 \hline
\end{tabular}
\label{dbarret_tab2}
\end{table*}

\begin{figure}
\includegraphics[height=0.3\textheight]{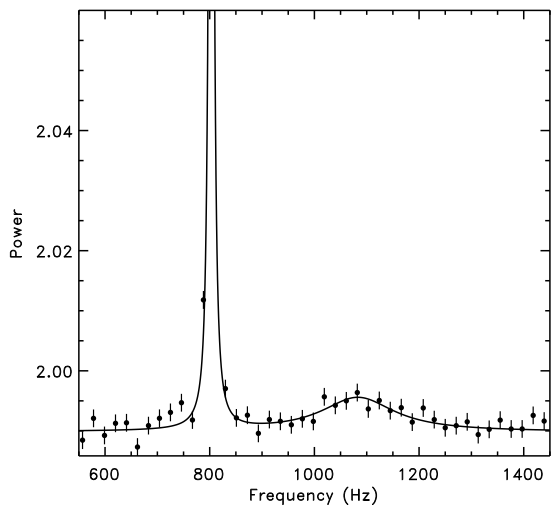}
\caption{The lower and upper kHz QPOs of Aql X-1, combining the data of Table 1 and 2. The upper QPO is detected at $3.2\sigma$ (the PDS has been linearly binned for illustrative purposes). }
\label{dbarret_fig4}
\end{figure}

Assuming that the QPOs detected are indeed lower kHz QPOs, one can
shift-and-add all of them to a reference frequency. By doing that for
all the observations listed in Table \ref{dbarret_tab1}, the second
strongest (and the only one) excess of the resulting PDS, is found
above the main peak at a frequency separation which is consistent
with half the spin frequency of the neutron star (275 Hz). Combining
the observations of Table \ref{dbarret_tab1} and those of Table
\ref{dbarret_tab2}, increases the significance of the detection above
$3 \sigma$. The results of the fitted QPOs in both cases are listed
in Table \ref{dbarret_tab3}. This, together with the fact that the
frequency separation is exactly in the expected range, gives us
strong confidence that our QPO detection is real. The two QPOs detected by combining the segments of Tables \ref{dbarret_tab1} and \ref{dbarret_tab2} are shown in Fig. \ref{dbarret_fig4}. We have searched
for an upper QPO 250-300 Hz above the lower QPO within individual
ObsIDs and found no significant (above $3\sigma$) QPOs (upper limit
ranging from $\sim 6$ to 10\% RMS, depending on the source count rate
for a QPO width of 100 Hz). Clearly our detection has been made
possible through the use of the shift and add technique and the
combination of more than $\sim 70$ kseconds of data.

\begin{table*}
 \centering
   \caption{Lower and upper kHz QPOs from Aql X-1. The left column indicates the origin of the data averaged. T$_{obs}$ is the cumulative integration time of all the PDS used to detect the upper QPO. $\nu_{\rm lower}$, $\bar{Q}_{\rm lower}$, RMS $_{\rm lower}$ are respectively the frequency, quality factor, and amplitude of the lower QPO. $\nu_{\rm upper}$, $\bar{Q}_{\rm upper}$, RMS$_{\rm upper}$ are the same parameters for the upper QPO. $\sigma_{\rm upper}$ is the significance of the upper QPO. $\Delta\nu$ is the frequency difference between the two QPOs. All errors are again computed such that $\Delta\chi^2=1$.}
 \begin{tabular}{@{}|lccccccccc|@{}}
\hline 
 & T$_{obs}$ & $\nu_{\rm lower}$ & $\bar{Q}_{\rm lower}$ & RMS $_{\rm lower}$ & $\nu_{\rm upper}$ & $\bar{Q}_{\rm upper}$ & RMS$_{\rm upper}$ &$\sigma_{\rm upper}$ & $\Delta\nu$ \\
 \hline
Table 1 & 69824&$795.45\pm0.04$&$129.32\pm2.25$&$7.26\pm0.05$&$1073.5\pm18.2$&$5.7\pm2.1$&$4.4\pm0.8$&2.6&$278.1\pm18.3$\\
Table 1 \& 2 & 86432&$803.09\pm0.05$&$120.44\pm2.22$&$6.60\pm0.04$&$1083.2\pm13.3$&$6.3\pm2.0$&$4.3\pm0.7$&3.2&$280.1\pm13.4$\\
  \hline
\end{tabular}
\label{dbarret_tab3}
\end{table*}

\section{Discussion}
We have studied the properties of the lower kHz QPOs from Aql X-1. 
We have fewer details than for other sources, but the behavior of
the lower QPOs is similar to that seen previously. More observations
of this source are needed to fully sample the quality factor versus
frequency diagram, in particular the high frequency part, where the
drop of coherence may be detected (around 900 Hz). The main result
of this paper is the detection for the first time of an upper kHz
QPO in this source, with an average frequency separation that is
consistent with half the spin frequency of the neutron star.

This result affords us a fresh opportunity to evaluate the 
relation between the spin frequency
and the QPO separation frequency in neutron star LMXBs.  It has long
been known that the separation frequency is not constant in a
given source, and indeed can change in rather complicated ways
(e.g., see the data for 4U~1608--52 in Figure~3 of M\'endez et
al. 1998).  As a result, no simple model can reproduce exactly
the observed behavior.  However, although the absolute goodness
of fit of simple models is therefore poor, it is possible to do a
statistical comparison between candidate models (e.g., through a
$\Delta\chi^2$ test), to determine which is closest to current
data and perhaps to provide guidance about the underlying physics.

To do this we compare six models.  ``Spin" is the most 
commonly discussed model, in which the separation is equal to the spin
frequency if $\nu_{\rm spin}<400$~Hz, but equal to half the spin
frequency otherwise.  ``Const" assumes a constant frequency
separation for all sources. ``Linear" applies the formula $\Delta\nu=390~{\rm
Hz}-0.2\nu_{\rm spin}$ from Yin et al. (2007).  ``Epicycle" is
the proposal (Stella \& Vietri 1998) that the upper peak frequency
is the orbital frequency at some radius and the lower peak is the
radial precession frequency at that same radius, meaning that the 
difference frequency is expected to change and to be equal to the
radial epicyclic frequency at the given radius (this therefore requires
an assumed mass for each source).  ``Power law" is
inspired by Psaltis, Belloni, \& van der Klis 1999: 
$\nu_{\rm upper}=\left(\nu_{\rm lower}/\nu_0\right)^p$, where
$\nu_0$ and $p$ are the same for all sources.  Finally,
``Ratio" is a model following Abramowicz et al. (2003), in which
the ratio between the upper and lower kHz QPO is fixed at the same
value for all sources.

We compare these models to the data available from the ten
sources listed by M\'endez \& Belloni (2007), as well as
Aql~X-1 and 4U~0614+091 (see footnote to Table~\ref{dbarret_tab4} 
for the list of sources and
primary references).  There are a total of 57 independent
measurements among these 12 sources, from which we compute
total $\chi^2$ values for the four models.  We note that
M\'endez \& Belloni (2007) suggest that the accretion-powered
millisecond pulsars XTE~J1807--294 and SAX~J1808--3658 should
be treated specially because some of their other frequency
properties appear offset by a factor of roughly 1.5 from
those of other sources.  We thus compute total $\chi^2$ values
omitting these sources, and also including these sources but
multiplying their frequency separations by a factor of 1.5,
to evaluate the robustness of the model comparisons.

Table~\ref{dbarret_tab4} shows the results.  The Stella \& Vietri
(1998) ``Epicycle" model appears at first to be competitive when the
millisecond pulsars are ignored or have their frequencies adjusted. 
This, however, is somewhat misleading: four sources (4U~1702--43,
IGR~J17191--2821, KS~1731--260, and SAX~J1750.8--2900) have only one
measurement each of a separation frequency, so it is possible to
pick a neutron star mass that fits the single data point perfectly
in those cases.

\begin{table*}
\caption{$\chi^2$ values for models of QPO frequency separation. 
$^{\rm a}$See text for description of models. $^{\rm b}$All available 
published data, for
XTE~J1807--294 (Linares et al. 2005), SAX~J1808.4--3658 (Wijnands
et al. 2003), 4U~1608--52 (M\'endez et al. 1998), 4U~1636--536
(Di Salvo et al. 2003), 4U~1702--43 (Strohmayer et al. 1998),
4U~1728--34 (M\'endez \& van der Klis 1999), 4U~1731--260
(Wijnands \& van der Klis 1997), IGR~J17191--2821
(Klein-Wolt et al. 2007), SAX~J1750.8--2900 (Kaaret et al. 2002),
4U~1915--05 (Boirin et al. 2000), Aql~X-1 (this work), and
4U~0614+091 (twin QPO frequencies taken from  Barret, Olive, Miller (2006) and spin frequency from Strohmayer, Markwardt \& Kuulkers (2007)).
$^{\rm c}$Same as ``Full data set" except that we removed
the data points due to the accretion-powered millisecond pulsars
XTE~J1807--294 and SAX~J1808.4--3658. $^{\rm d}$Same as ``Full data set" 
except that the frequency
separations for XTE~J1807--294 and SAX~J1808.4--3658 were multiplied
by 1.5, following M\'endez and Belloni (2007).}
  \begin{tabular}{@{}|cccc|@{}}
\hline
Model$^{\rm a}$ & $\chi^2$ - Full data set$^{\rm b}$ & 
$\chi^2$ -Reduced data set$^{\rm c}$ & $\chi^2$ - 
Modified data set $^{\rm d}$ \\
\hline
Spin & 1628 & 1616 & 2200 \\
Const & 2994 & 2368 & 2391 \\
Linear & 2566 & 1666 & 1736 \\
Epicycle & 1742 & 1275 & 1305 \\
Power law & 1942 & 1194 & 1420 \\
Ratio & 17945 & 17623 & 19388 \\
\hline
\end{tabular}
\label{dbarret_tab4}
\end{table*}

From these data we can draw a few conclusions:
\begin{itemize}
\item All simple models fail badly in a statistical sense.  There is
clearly unmodeled complexity to these systems.  For an example of how
such complexity might affect the frequencies in the ``Spin" model, see
Lamb \& Miller (2001).
\item Of the models considered, the one assuming a constant ratio is
overwhelmingly the worst, for any of our data sets.  The next worst
in all cases (but by a much smaller margin) is the one assuming a
constant difference frequency for all systems.  Other models are
preferred by the data.
\item The treatment of data from the accretion-powered millisecond
pulsars XTE~J1807--294 and SAX~J1808.4--3658 has a significant effect
on the comparison between the remaining models.  With the data as is,
the standard ``Spin" model does best. If the frequency differences
for just these sources are multiplied by a factor of 1.5, as
advocated by M\'endez \& Belloni (2007), then the ``Epicycle" and
``Power law" models do best.  It is not clear how significant this
is; we note, for example, that for any particular model, if the two
worst-fit sources are eliminated, the fit becomes much better in all
cases.

\end{itemize}

We conclude that although the separation frequency is clearly a
complex quantity, the standard model fits the data at least
comparably well to similarly simple models, in addition to emerging
from generally plausible input physics.  It is therefore still a
viable hypothesis that the spin frequency affects the kHz QPOs seen
from neutron star low-mass X-ray binaries. On the other hand, the
generally bad fits of all models and the possibly important role of
individual sources both raise the unpalatable but real possibility
that there are multiple mechanisms that can produce kHz QPOs in
neutron-star LMXBs.

\section{Conclusions}


We have shown that the properties of the QPOs detected from Aql X-1
so far are consistent with those seen from similar systems, in
particular the quality factor of the lower QPOs and the frequency
separation between the lower and upper peaks, which we have
measured for the first time to be close to half the spin frequency
of the neutron star. It would be worth following up the lower QPOs
closer to the saturation frequency (at 900 Hz), with adequate
sensitivity to estimate the quality factor of the QPOs, and
determine whether it drops as in other sources. This may become
possible with RXTE during the next outburst of this very active 
transient.

\section{Acknowledgements}

MCM was supported in part by NSF grant AST0708424.  This research has
made use of data obtained from the High Energy Astrophysics
Science Archive Research Center (HEASARC), provided by NASA's
Goddard Space Flight Center. We are grateful to Jean-Francois Olive 
for useful discussions during the preparation of this paper. We thank the referee for comments that helped up to improve the presentation of the results reported in this paper.

\end{document}